\documentclass{aa}
\usepackage{array}

\def\dg{^{\circ}}
\newcommand{\tabsize}{\fontsize{8}{2}\selectfont}
\newcolumntype{R}{>{\tabsize}r}
\newcolumntype{L}{>{\tabsize}l}
\newcolumntype{C}{>{\tabsize}c}
\newcolumntype{P}{@{\tabsize \, (}R@{\tabsize )\quad}}

\newcommand{\gppr}{\stackrel{>}{\scriptstyle \sim}}
\newcommand{\gappr}{\raisebox{-0.4ex}{$\gppr$}}
\newcommand{\lppr}{\stackrel{<}{\scriptstyle \sim}}
\newcommand{\lappr}{\raisebox{-0.4ex}{$\lppr$}}
\def\Lalpha{\mbox{Lyman\,$\alpha$}}
\def\Grw{\mbox{Grw~$+70\dg 8247$}}
\def\He1211{\mbox{HE\,1211-1707}}
\usepackage{times}
\usepackage{graphics}
\usepackage{epsfig}
\begin{document}
\thesaurus {08(08.23.1, 08.13.1, 08.09.2:GD229}
\title{{\em Letter to the Editor} \\
  {Evidence for helium in the magnetic white dwarf GD\,229}}
\author {S. Jordan\inst{1}  
\and P. Schmelcher\inst{2}
\and W. Becken \inst{2}
\and W. Schweizer \inst{3}
}
\institute{Institut f\"ur Theoretische
 Physik und Astrophysik, D-24098 Kiel, Germany
\and
Institut f\"ur Physikalische Chemie, Im Neuenheimer Feld 253, D-69120 Heidelberg,
Germany
\and
Institut f\"ur Astronomie und Astrophysik, Auf der Morgenstelle 10, D-72076 
T\"ubingen, Germany}

\offprints{S.~Jordan, jordan@astrophysik.uni-kiel.de}
\date{received June 15, 1998; accepted June 18, 1998}
\titlerunning{Solution of the mystery of GD\,229}
\authorrunning{S. Jordan et al.}
\maketitle
\begin{abstract}
The nature of the strong absorption features in  the  white dwarf
GD\,229 has been a real mystery ever since it was found to be magnetic
in 1974. All attempts to explain the spectrum by line components of
hydrogen failed. With the first sets of newly calculated line data for
\ion{He}{I} in a strong magnetic field we could identify most of the
absorption structures in the spectrum with stationary line components
in a  range of magnetic fields  between 300 and 700\,MG. This is much lower
than previously
speculated and is comparable to the highest fields found in hydrogen rich
magnetic white dwarfs. The reason for the large number of spectral features
in GD\,229 is the extreme accumulation  of stationary
components of transitions in He I
in a narrow interval of field strengths. 

\end{abstract}

\section{Introduction}
   Magnetic fields from about 40 kG to 1 GG have been detected in about
    about 50 (2\%) of the 2100 known white dwarfs (McCook \&\ Sion 1998).
    A list
    of all currently known magnetic white dwarfs is found in Jordan (1997).
    Magnetic fields in the low field range ($\lappr 20$\,MG) in both 
    hydrogen and
    helium rich magnetic white dwarfs can be detected by a relatively simple
    line pattern. Hydrogen line components have been identified in many
    objects, but until recently, helium was detected unambiguously only in
    the magnetic white dwarf Feige\,7. Using \ion{He}{I} line data calculated by
    Kemic (1974), Achilleos et al. (1992) found that the spectrum could be
    well reproduced with a dipole of strength 35\,MG displaced by 0.15 white
    dwarf radii from the stellar center. Since the Kemic data were calculated
    by regarding the magnetic field as a small perturbation to the Coulomb
    interaction and are only valid up to about 20\,MG, the Kemic tables had to
    be extrapolated for the Feige 7 analysis. The only other definite helium
    rich objects with fields of about 15-20\,MG were discovered by Reimers et
    al. (1998) in the course of the Hamburg ESO survey.

Magnetic white dwarfs with  field strengths $\gappr 100$\,MG
containing hydrogen could be interpreted ever since the numerical calculations of
energy level shifts and transition
probabilities for bound-bound transitions by groups in T\"ubingen and Baton
Rouge (Forster et al. 1984; R\"osner et al. 1984; Henry and O'Connell 1984,
1985). With the help of these data 
Greenstein 1984 and  Greenstein et al. 1985  could identify the
hitherto unidentified features in \Grw\ with stationary components of
hydrogen in fields between about 150 and 500\,MG.

Since  the magnetic field on the surface of a white dwarf is not
homogeneous but e.g. better described by a magnetic dipole, the variation of the
field strength from the pole to the equator (a factor of two for a pure dipole
field) smears out most of the absorption lines at larger magnetic field
strengths. However, a few of the line components become
stationary, i.e. their wavelengths go through maxima or minima as functions of
the magnetic field strength. These stationary components are visible in the   
spectra of magnetic white dwarfs despite a considerable variation of the field  
strengths.

About 80\,\%\ of all known white dwarfs have nearly pure hydrogen
atmospheres (spectral type DA). However, since most of the remaining stars
have helium 
rich atmospheres, we would also expect a significant
number of magnetic white dwarfs to belong to the spectral type DB in which
\ion{He}{I} lines are observed.
Therefore it could be expected that  the few magnetic white dwarfs
with unidentified spectral features, that cannot be explained by the
line data for hydrogen,  are helium rich.
The most  famous example
 is  GD\,229, where Swedlund et al. (1974), Greenstein et al. (1974), 
 Landstreet \&\ Angel (1974),
Liebert (1976), Greenstein \&\ Boksenberg (1978), and Schmidt et al.
(1996) found strong absorption features in the optical, infrared, and
UV. Angel (1979) already proposed that the absorption bands 
in    this star are due to stationary components of hydrogen or helium.

The basic difficulty in calculating energy levels and oscillator
strengths for arbitrary field strengths lies in the fact that the
Coulomb potential has a spherical symmetry whereas the magnetic
field induces a cylindrical symmetry. This prevents a separation
of variables and together with the nonlinear character of the
interactions makes even the numerical solution a difficult problem.
In particular the intermediate regime ($0.01 < \gamma
= B / 2.35 \cdot 10^{9} G < 1$), in which the majority of magnetic
white dwarfs is found, is very complicated, since neither the magnetic
nor the Coulomb interaction dominates and we therefore encounter
a so-called nonperturbative regime where none of the interaction
terms can be treated by perturbation theory. In the case of the
two electron system \ion{He}{I} the situation is even more difficult, 
since the number of degrees of freedom is enhanced significantly
and the electron-electron interaction introduces a third
competitive force which makes \ion{He}{I} a complex system with a rich
variability of the spectrum with changing field strength.
The particular challenge of a comparison of atomic data
with observational ones lies in the fact that the energies
of many excited states have to be known with a high relative
accuracy (usually $\le 10^{-4}$) and for a large number of 
field strengths in order to identify the stationary transitions.

Until this year the status of electronic structure
calculations on \ion{He}{I} was such (see Braun et al.1998, Jones et al. 1997,
 Ruder et al.1994 and references
therein) that a conclusive comparison was
impossible and the occurence of \ion{He}{I} in the atmosphere
of GD\,229 had to be considered as a pure speculation as it
was the case with respect to hydrogen in \Grw\ until
1984.

Therefore, several alternative explanations have been proposed.
Engelhardt \&\ Bues (1995) have tried to explain the
regular almost periodical structure of the GD\,229 spectrum by quasi-Landau
resonances  of hydrogen in a magnetic field of 2.5\,GG. \"Ostreicher
et al. (1987) speculated that two of the features could be due to
intersections of  hydrogen components in a field of about 25-60\,MG.

Now, the calculations performed in Heidelberg (Becken \& Schmelcher 1998)
have closed that gap and
for the first time precise data for a large number of energy states have
become available. In this paper we will show that most of the 
absorption features in the optical and UV spectrum of GD\,229 can
be identified as stationary line transitions of \ion{He}{I}.

\section{Helium in strong magnetic field}
The fixed-nucleus electronic
Hamiltonian of the helium atom in a strong magnetic field oriented along the
$z-$axis reads as follows 
\begin{eqnarray}
\cal{H} = & \sum_{i=1,2} \left(\frac{1}{2m} {\bf{p}}_i^2 +\frac{e}{2m}
 B L_{z_i} \right.
+ \frac{e^2}{8m} B^2 (x_i^2 + y_i^2) \nonumber \\
&\left. - \frac{2e^2}{|{\bf{r}}_i|} 
 + \frac{e}{m} B S_{z_i} \right) + \frac{e^2}{|{\bf{r}}_1-{\bf{r}}_2|} 
\end{eqnarray}
where the sum includes the one particle terms in the order:
field-free kinetic, orbital Zeeman, diamagnetic, Coulomb attraction and
spin Zeeman energies. The last term represents the repulsive electron-electron
interaction. Effects of the finite nuclear mass can easily be taken into account
by using the corresponding scaling relations (Becken \&\ Schmelcher 1998).
According to its
symmetries the Hamiltonian (1) possesses four independent conserved 
quantities:
the total spin ${\bf{S}}^2$, its $z-$component $S_z$, the $z-$component $L_z$ of the
total spatial electronic angular momentum and the total spatial $z-$parity $\Pi_z$
(parity is a combined symmetry of the previous ones) which will be used to label
the electronic states in the form $n^{2S+1}M^{(-1)^{\Pi_z}}$ where $n$ specifies
the degree of excitation within a given symmetry subspace. 
Calculations are performed separately for each subspace with given quantum numbers.
Since the spin part of the eigenfunctions of the Hamiltonian (1) are the usual
spin singlet and triplet functions we address in the following exclusively the
method of investigation for the spatial part of the wave functions.
 
The key ingredient of our method for the ab initio calculation of the electronic
structure of atoms in strong magnetic fields is a basis set of one particle functions
which takes on the following appearance
\begin{equation}
\Phi_i (\rho,\phi,z) = \rho^{n_{\rho_i}} z^{n_{z_i}} \mbox{exp}\left(-\alpha_i\rho^2
-\beta_iz^2\right) \mbox{exp}\left(i m_i \phi\right)
\end{equation}
where $n_{\rho_i}=|m_i|+2k_i, n_{z_i}= \pi_{z_i} + 2l_i$ with $k_i,l_i = 0,1,2,...$
$\{\alpha_i,\beta_i\}$ are nonlinear variational parameters; $\pi_{z_i}$
is the $z$ parity of the one particle function.
The basis set (2) has its precursor in a more general one 
(Schmelcher \&\ Cederbaum 1988)  which has
successfully been applied for the investigation of the electronic structure of
small molecules in strong magnetic fields (Kappes \&\ Schmelcher 1996).
 
Due to the parameters $\{\alpha_i,\beta_i\}$ our basis set is extremely flexible
which represents a major advantage in the presence of an external magnetic field.
This allows us to accurately describe the changes in the electronic wave functions
with changing field strengths. The values of the variational parameters
 $\{\alpha_i(B), \beta_i(B)\}$ are determined by a nonlinear optimization procedure which is performed
for the spectrum of both hydrogen and He II in a magnetic field.
Typically 150 one particle functions of type (2) are optimized in order to describe
15 hydrogenic states with a high accuracy.
The optimization procedure has to be accomplished for each field
strength separately (see Becken \&\ Schmelcher 1998 and references therein).
 
The basic idea for solving the time-independent Schr\"odinger equation belonging
to the Hamiltonian (1) is then to construct out of the above optimized one
particle functions a set of spatial two electron configurations $\{\psi_i\}$
being symmetric (spin singlet) or antisymmetric (spin triplet) with respect to
the interchange of the electronic coordinates and respecting the above-discussed
symmetries. We represent the exact  wave functions $\Psi_j$ by a
linear combination of two electron configurations
$\Psi_j = \sum_{i} c_{ji} \psi_i({\bf{r}}_1,{\bf{r}}_2)$.
By virtue of the variational principle we  arrive at a generalized eigenvalue
problem whose eigenvalues and eigenvectors are excellent approximations
to the exact eigenenergies and wave functions. We remark that all matrix elements
of the Hamiltonian (1) with respect to the basis afunctions (2) can be evaluated
 analytically resulting in a complicated series of higher
 transcendental functions
whose numerically stable and efficient implementation 
required a number of different techniques.
 
The typical number of two-electron configurations $\{\psi_i\}$, i.e. the dimension
of the full Hamiltonian matrix to be diagonalized, is of the order of 2000-4000
depending on the symmetries. 
On a powerful workstation 10 - 20 h CPU are necessary to build up
and diagonalize the Hamiltonian matrix for each field strength
and symmetry subspace.
Extensive calculations have been performed in the
whole range $0\le\gamma=B/2.35\cdot 10^9{\rm G}\le 100$,
for 30  electronic states
the energy could be determined to an accuracy of at least $10^{-4}$.
For vanishing magnetic quantum number $M=0$ many excited singlet and triplet electronic states
with positive and negative $z-$parity have been determined.
For nonzero angular momentum $M=+1,-1$ both singlet and triplet states
have been studied for positive $z-$parity. This gives us insight into an essential
part of the spectrum of the He I atom with varying field strength and allows
us in particular to decide whether He I can be present in the atmosphere of
magnetic white dwarfs through an extraction of the stationary components of the
transitions in the corresponding regime of field strengths.

\section{Interpretation of the spectrum}
The strongest evidence that GD\,229 does not contain hydrogen came 
from a HST spectrum taken by Schmidt et al. (1996):
There is no evidence for components of \Lalpha\ which do not
vary very much over a very large range of magnetic field strengths.
The $1s_0\rightarrow2p_{-1}$ transition has a maximum at 1341\,\AA\
and should always be visible even at extremely high field if
significant amount of hydrogen were present.

For the first time it is now possible to compare the spectral features
of GD\,229 directly with the newly calculated wavelength tables for the
line components of \ion{He}{I} in a strong magnetic
field  (Becken\,\& Schmelcher 1998).
Since oscillator strengths are not yet available
we could not directly compare the spectrum with theoretical models
(see e.g. Jordan 1992).  Therefore we  compared the positions 
of stationary line components with the observed absorption bands.
In total about 150 line components have been calculated of which
about 40 show  stationary behaviour at different field strengths.

\begin{figure}
  \begin{center}
    \epsfig{file= 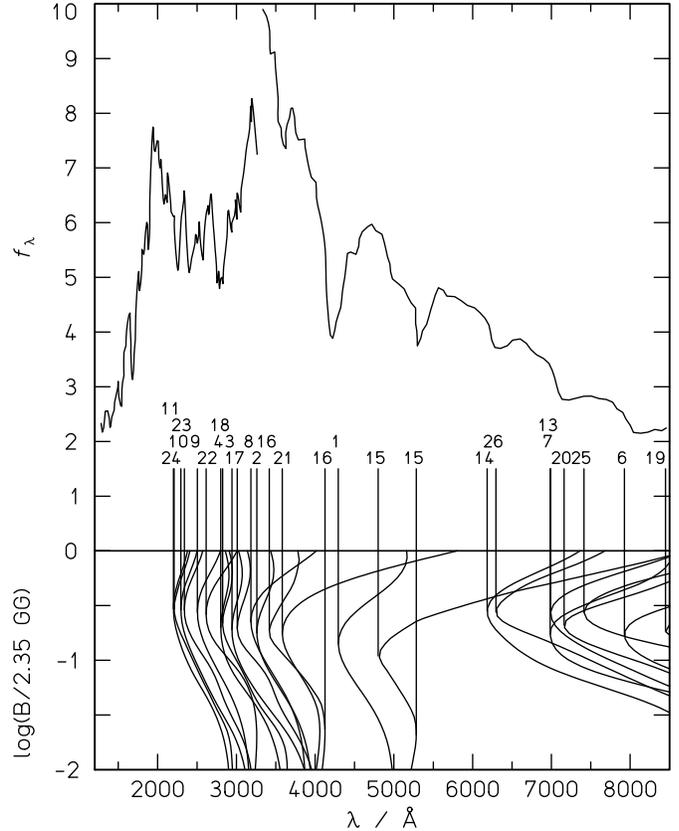,
     clip=, width=8.8cm,
      bbllx=49.8, bblly=42, bburx=539.2, bbury=652.5}
  \end{center}
  \caption{
The spectrum of \object{GD~229} (Schmidt et al. 1996) is compared with
the positions of \ion{He}{I} components which are stationary in the
plotted range  of field strengths. The stationary wavelengths are marked
and labeled with a number corresponding to the identification in 
Tab.\,\ref{stationaries}. Note that additional stationary line components 
which are not yet theoretically calculated may contribute
 }
\label{gdfig} 
\end{figure}

If one looks at the helium line data up to $\gamma=100$ (corresponding
to $235$\,GG) it is rather striking to see that the majority (26)
of all stationary 
components 
lies in  the range  $300\lappr B/{\rm MG} \lappr 700$
( see Fig.\,\ref{gdfig}; note that the wavelengths were determined
 by spline interpolation
in a still relatively crude grid). So far only 10 grid points exist for the 
range of magnetic
field strengths plotted in Fig.\,\ref{gdfig}.

In practically all cases
the minimum wavelength of the  large number of features seen in the optical, 
IR, and UV
spectrum of GD\,229  can be attributed to the wavelength 
minima  of the helium
components (listed in Table\,\ref{stationaries}).
This explains
why the spectrum of GD\,229 has such an extremely large number of
absorption features.

Moreover, there are no clear contradictions between the predicted
positions of absorptions and the observations. The only shortcoming
of the current analysis is the fact that the strongest features
at $4000-4200$\,\AA\ and at about $5280$\,\AA\  can only be attributed to two
line transitions
($2^1 0^+\rightarrow 2^1 0^-$, $2^1 0^+\rightarrow 2^1 (-1)^+$). 
Since the identification at other wavelengths is 
convincing we believe that additional stationary line components
are hidden in the subspaces of bound-bound transitions that have
not been calculated yet (four subspaces $^1(\pm 1)^-$,  $^3(\pm 1)^-$,
and  transitions with magnetic quantum numbers $|m|>1$).

In principle the  wavelength maxima (also shown in the figure) of the
 $2^1 0^+\rightarrow 3^1 (-1)^-$ and  $2^1 0^+\rightarrow 2^1 (-1)^-$
transitions  would fit well to the position of the two strongest absorption.
However, the corresponding field strengths of about 50\,MG would 
be much too low compared to the result found for the other features.
Such a large spread of magnetic field strengths is also rather improbable since
the observed features are relatively sharp.

An alternative interpretation would be that hydrogen is additionally
present in the spectrum. The strong feature at 4124\,\AA\ in the
spectrum of \Grw\ (which has a polar field strength of 320\,MG, e.g.
Jordan 1992) 
caused by the hydrogen transition $2s_0\rightarrow 4f_0$
could also contribute to the strongest absorption  feature 
in GD\,229. However, in this case we would  expect several other 
absorption troughs, e.g. at 5830\,\AA\ ($2s_0\rightarrow 3p_0$) which
are not present in GD\,229. This is also compatible with the finding by
Schmidt  et al. (1996)  that no Lyman\,$\alpha$ components are present
in the UV spectrum.

\begin{table}[htbp]
  \caption[]{ Components of \ion{He}{I} which are
stationary at $300<B/{\rm MG}<700$ and additionally two maxima
at $\approx 50$\,MG. Note that the wavelengths (and corresponding
field strenghts) of the minima and maxima are interpolated in a relatively
crude grid. The error range in $\lambda$ was estimated from interpolations of different
orders
}
  \begin{center}
    \begin{tabular}{RCCR}\hline
No. &   Component & Zero field trans. & \multicolumn{1}{c}{($\lambda$/\AA) $|$($B$/MG)}  \\\hline
1  &    $2^1 0^+\rightarrow 2^1 0^-$ & $2^1 S_0 \rightarrow 3^1 P_0$ & 4294 $\pm$8    $|$ 338 min \\
2   &    $2^1 0^+\rightarrow 3^1 0^-$ & $2^1 S_0 \rightarrow 4^1 F_0$ & 3260 $\pm$ 6   $|$  425 min \\
3   &    $2^1 0^+\rightarrow 4^1 0^-$ & $2^1 S_0 \rightarrow 4^1 P_0$ & 2943 $\pm$ 10  $|$  455 min \\
4   &    $2^1 0^+\rightarrow 5^1 0^-$ & $2^1 S_0 \rightarrow 5^1 F_0$ & 2802 $\pm$  5  $|$  477 min \\
5  &    $3^1 0^+\rightarrow 3^1 0^-$ & $3^1 S_0 \rightarrow 4^1 F_0$ &10692 $\pm$  40 $|$  379 min \\
6  &    $3^1 0^+\rightarrow 4^1 0^-$ & $3^1 S_0 \rightarrow 4^1 P_0$ & 7927 $\pm$ 40  $|$  406 min \\
7  &    $3^1 0^+\rightarrow 5^1 0^-$ & $3^1 S_0 \rightarrow 5^1 F_0$ & 6984 $\pm$ 40  $|$  415 min \\
\hline
8   &    $1^3 0^+\rightarrow 2^3 0^-$ & $2^3 S_0 \rightarrow 3^3 P_0$ & 3183 $\pm$ 6   $|$  535 min \\
9   &    $1^3 0^+\rightarrow 3^3 0^-$ & $2^3 S_0 \rightarrow 4^3 P_0$ & 2504 $\pm$ 3   $|$  658 min \\
10   &    $1^3 0^+\rightarrow 4^3 0^-$ & $2^3 S_0 \rightarrow 4^3 F_0$ & 2295 $\pm$ 3   $|$  689 min \\
11  &    $1^3 0^+\rightarrow 5^3 0^-$ & $2^3 S_0 \rightarrow 5^3 P_0$ & 2200 $\pm$ 5   $|$  705 min \\
12  &    $2^3 0^+\rightarrow 3^3 0^-$ & $3^3 S_0 \rightarrow 4^3 P_0$ & 9350 $\pm$ 6   $|$  573 min \\
13  &    $2^3 0^+\rightarrow 4^3 0^-$ & $3^3 S_0 \rightarrow 4^3 F_0$ & 6990 $\pm$ 10  $|$  643 min \\
14  &    $2^3 0^+\rightarrow 5^3 0^-$ & $3^3 S_0 \rightarrow 5^3 P_0$ & 6184 $\pm$ 10  $|$  674 min \\
\hline
15  &    $2^1 0^+\rightarrow 2^1 (-1)^+$ & $2^1 S_0 \rightarrow 3^1 P_{-1}$ &5285 $\pm$ 3  $|$   50 max \\
15  &    $2^1 0^+\rightarrow 2^1 (-1)^+$ & $2^1 S_0 \rightarrow 3^1 P_{-1}$ &4812 $\pm$ 30 $|$  258 min \\
16  &    $2^1 0^+\rightarrow 3^1 (-1)^+$ & $2^1 S_0 \rightarrow 4^1 F_{-1}$ &4125 $\pm$ 3  $|$   56 max \\
16  &    $2^1 0^+\rightarrow 3^1 (-1)^+$ & $2^1 S_0 \rightarrow 4^1 F_{-1}$ &3417 $\pm$ 10 $|$  429 min \\
17  &    $2^1 0^+\rightarrow 4^1 (-1)^+$ & $2^1 S_0 \rightarrow 4^1 P_{-1}$ &3012 $\pm$ 10 $|$  466 min \\
18  &    $2^1 0^+\rightarrow 5^1 (-1)^+$ & $2^1 S_0 \rightarrow 5^1 F_{-1}$ &2825 $\pm$ 3  $|$  523 min \\
19  &    $3^1 0^+\rightarrow 4^1 (-1)^+$ & $3^1 S_0 \rightarrow 4^1 P_{-1}$ &8449 $\pm$ 150$|$  445 min \\
20  &    $3^1 0^+\rightarrow 5^1 (-1)^+$ & $3^1 S_0 \rightarrow 5^1 F_{-1}$ &7160 $\pm$ 20 $|$  499 min \\
\hline
21  &   $1^3 0^+\rightarrow 2^3 (-1)^+$ & $2^3 S_0 \rightarrow 3^3 P_{-1}$ &3582 $\pm$ 2   $|$  415 min \\
22  &   $1^3 0^+\rightarrow 3^3 (-1)^+$ & $2^3 S_0 \rightarrow 4^3 P_{-1}$ &2615 $\pm$ 5   $|$  629 min \\
23  &   $1^3 0^+\rightarrow 4^3 (-1)^+$ & $2^3 S_0 \rightarrow 4^3 F_{-1}$ &2338 $\pm$ 5   $|$  689 min \\
24  &   $1^3 0^+\rightarrow 5^3 (-1)^+$ & $2^3 S_0 \rightarrow 5^3 P_{-1}$ &2208 $\pm$ 3   $|$  689 min \\
25  &   $2^3 0^+\rightarrow 4^3 (-1)^+$ & $3^3 S_0 \rightarrow 4^3 F_{-1}$ &7413 $\pm$ 40  $|$  658 min \\
26  &   $2^3 0^+\rightarrow 5^3 (-1)^+$ & $3^3 S_0 \rightarrow 5^3 P_{-1}$ &6296 $\pm$ 40  $|$  658 min \\
    \end{tabular}
  \end{center}
\label{stationaries}
\end{table}

%

\section{Discussion and future prospects}
For the first time we have been able to prove that GD\,229 is indeed a
helium rich (DB)  white dwarf at a rather high magnetic field strength.
>From the magnetic field strengths corresponding to the extrema of the
wavelengths of the line components we conclude that the range of magnetic
fields is about 300-700\,MG, so that it would be compatible with a
dipole field with a polar field strength between 600 and 700\,MG.
With our analysis we could reject previous ideas 
 that the atmosphere is hydrogen rich.  

In the near future the rest of the line data for \ion{He}{I} will be
calculated. We expect that some of the missing line
components will have wavelength extrema near the strong features
at about 4100 and 5300\,\AA.

A detailed modelling of the spectrum and polarization of GD\,229
will be possible as soon as  oscillator strengths become  available;
this will enable us to determine the magnetic field structure which
is impossible by our  wavelength analysis.

With the new line data for \ion{He}{I} we will also try to interpret other
magnetic white dwarfs with unexplained features: e.g. 
\He1211 (Reimers et al. 1994, Jordan 1997), and four other newly discovered
objects from the Hamburg ESO Survey (Reimers et al. 1998).

Another good candidate for such an analysis is 
LB\,11146B for which a convincing analysis of the spectrum
with hydrogen line components arrived at an approximate
field strength of about  
670\,MG (Liebert et al. 1993, Glenn et al. 1994).
Additionally a very strong and broad  feature at about 
5600\,\AA\ was   present in the spectrum which cannot be due to
hydrogen. Therefore Glenn et al. speculated that helium might be 
responsible. However, in our current set of line data no 
stationary line component of \ion{He}{I} was found at 
about 670\,MG.

{\em Acknowledgements.} The Deutsche Studienstiftung and the Deutsche
Forschungsgemeinschaft (Schm 885/5-1 \& Ko 735/7-1)
are gratefully acknowledged for financial
support. We thank Susanne Friedrich for the careful reading  of the 
manuscript and the referee Richard Allen for valuable comments.

\end{document}